\documentclass[conference]{IEEEtran}
\makeatletter
\def\ps@headings{%
\def\@oddhead{\mbox{}\scriptsize\rightmark \hfil \thepage}%
\def\@evenhead{\scriptsize\thepage \hfil \leftmark\mbox{}}%
\def\@oddfoot{}%
\def\@evenfoot{}}
\makeatother
\IEEEoverridecommandlockouts
\usepackage{cite}

\ifCLASSINFOpdf
\else
\fi
\usepackage{float}
\usepackage{subfigure}
\usepackage[cmex10]{amsmath}
\usepackage{amssymb}
\usepackage{graphicx}
\usepackage{color}
\usepackage{algorithmic}
\usepackage{algorithm}
\usepackage{amsthm}
\DeclareMathOperator*{\argmax}{arg\,max}
\DeclareMathOperator*{\subjectto}{subject\,to}

\begin{document}
\title{Backpressure with Adaptive Redundancy \\ (BWAR)
\thanks{This material is supported in part by the following: NSF grant 1049541, the
Network Science Collaborative Technology Alliance sponsored
by the U.S. Army Research Laboratory W911NF-09-2-0053, and King Saud University.}
}

\author{\IEEEauthorblockN{Majed Alresaini}
\IEEEauthorblockA{alresain AT usc . edu}
\and
\IEEEauthorblockN{Maheswaran Sathiamoorthy}
\IEEEauthorblockA{msathiam AT usc . edu}
\and
\IEEEauthorblockN{Bhaskar Krishnamachari}
\IEEEauthorblockA{bkrishna AT usc . edu}

\and
\IEEEauthorblockN{Michael J. Neely}
\IEEEauthorblockA{mjneely AT usc . edu}

}

\maketitle

\begin{abstract}
Backpressure scheduling and routing, in which packets are preferentially transmitted over links with high queue differentials, offers the promise of throughput-optimal operation for a wide range of communication networks. However, when the traffic load is low, due to the corresponding low queue
occupancy, backpressure scheduling/routing experiences long delays. This is particularly of concern in intermittent encounter-based mobile networks which are already delay-limited due to the sparse and highly dynamic network connectivity. While state of the art mechanisms for such networks have proposed the use of
redundant transmissions to improve delay, they do not work well when the traffic load is high. We propose in this paper a novel hybrid approach that we refer to as backpressure with adaptive redundancy (BWAR), which provides the best of both worlds. This approach is highly robust and distributed and does not require any prior knowledge of network load conditions. We evaluate BWAR through both mathematical analysis and simulations based on cell-partitioned model. We prove theoretically that BWAR does not perform worse than traditional backpressure in terms of the maximum throughput, while yielding a better delay bound. The simulations confirm that BWAR outperforms traditional backpressure at low load, while outperforming a state of the art encounter-routing scheme (Spray and Wait) at high load.
\end{abstract}

\section{Introduction}

Queue-differential backpressure scheduling and routing was shown by Tassiulas and Ephremides to be throughput optimal in terms of being able to stabilize the network under any feasible traffic rate vector~\cite{182479}. Additional research has extended the original result to show that backpressure techniques can be combined with utility optimization, resulting in simple, throughput-optimal, cross-layer network protocols for all kinds of networks~\cite{Georgiadis:2006:RAC:1166401.1166402,Neely2009862,4067770, 5935084, Bui09Infocom}. Recently, some of these techniques have been translated to practically implemented routing and rate-control protocols for wireless networks~\cite{Warrier09diffq:practical, BRCP-ITA, Avinash-JSAC, BCP, Ryu:2010:BRR:1859995.1860037}.

The basic idea of backpressure mechanisms is to prioritize transmissions over links that have the highest queue differentials. Backpressure effectively makes packets flow through the network as though pulled by gravity towards the destination, which has the smallest queue size of 0. Under high traffic conditions, this works very well, and backpressure is able to fully utilize the available network resources in a highly dynamic fashion. Under low traffic conditions, however, because many other nodes may also have a small or 0 queue size, there is inefficiency in terms of an increase in delay, as packets may loop or take a long time to make their way to the destination.

In this paper, we focus primarily on intermittently connected networks, such as encounter-based mobile networks (sometimes also referred to as delay or disruption tolerant networks (DTN)). In such networks, conventional path-discovery-based MANET routing techniques like AODV~\cite{perkins2002ad} and DSR~\cite{johnson1996dynamic} are not feasible because the network may not form a single connected partition at any time, and thus a full path may never exist between the source and the destination.  Instead, it is necessary to use store-and-forward type protocols that can handle the underlying mobility. A backpressure based routing scheme can be easily implemented in such a network, with the decision of what information to exchange being made between each pair of nodes based on their queue differentials whenever they encounter each other. However the above-mentioned delay inefficiency of the backpressure mechanism at low traffic loads is further exacerbated in such networks, because they are already delay-limited due to sparse network connectivity.

In the literature on intermittently connected networks, there are several proposed schemes for store-and-forward based routing, such as~\cite{4430783, Spyropoulos:2008:ERI:1373452.1373459, spraywait, Dubois-ferriere03agematters:, 1642727, Bai07impactof}. Some of these, such as Spray and Wait, advocate the use of redundant transmissions, to make additional copies of the communicated information in the network. The replication of the content makes it faster for the destination to access a copy. However, as the additional replication always increases the network load, these protocols, which are not throughput-optimal to begin with, suffer additional congestion.

In this paper, we propose a novel hybrid approach, an \emph{adaptive redundancy} technique for backpressure routing, that yields the benefits of replication to reduce delay under low load conditions, while at the same time preserving the performance and benefits of traditional backpressure routing under high traffic conditions. This technique, which we refer to as backpressure with adaptive redundancy (BWAR), essentially creates copies of packets in a new duplicate buffer upon an encounter, when the transmitter's queue occupancy is low. These duplicate packets are transmitted only when the original queue is empty. This mechanism can dramatically improve delay of backpressure during low load conditions due to two reasons: (1) due to the existence of multiple copies of the same packets at multiple nodes, the destination is more likely to encounter a massage intended for it. (2) this way, the algorithm builds up gradients towards the destinations faster and reduces packet looping.
The additional transmissions incurred by BWAR due to the duplicates utilize available slots which would otherwise go idle, in order to reduce the delay. Particularly for networks that are not energy-limited, this offers a more efficient way to utilize the available bandwidth during low load conditions. In order to minimize the storage resource utilization of duplicate packets, ideally, these duplicate packets should be removed from the network whenever a copy is delivered to the destination. Since this may be difficult to implement (except in some kinds of networks with a separate control plane), we also propose and evaluate a practical timeout mechanism for automatic duplicate removal. Under high load conditions, because queues are rarely empty, duplicates are rarely created, and BWAR effectively reverts to traditional backpressure and inherits its throughput optimality property. By design, BWAR is highly robust and distributed and does not require prior knowledge of locations, mobility patterns, and load conditions.

The following are the key contributions of this work:

\begin{itemize}

\item We propose BWAR, a new adaptive redundancy technique for backpressure scheduling/routing in intermittently connected networks. And we present a timeout mechanism for duplicate removal, which allows BWAR to be easily implemented in practice.

\item We develop an analytical model of BWAR, and prove theoretically that it yields a smaller upper bound on the average queue size (and hence the average delay) than traditional backpressure, while retaining throughput optimality.

\item Through simulations using an idealized cell-partition mobility model, we quantify the benefits from using BWAR. Specifically, we show that it outperforms both traditional backpressure and Spray \& Wait~\cite{spraywait}, a state of the art DTN/ICN routing mechanism.

\end{itemize}

The rest of the paper is organized as follows. In section~\ref{section:BWAR}, we introduce and describe BWAR. In section~\ref{section:reviewofbackpressure}, we review the theory behind traditional backpressure scheduling and routing.  We show in section~\ref{section:bwarqueueingdynamics} the queue dynamics for BWAR and how it can improve the delay theoretically. In section~\ref{section:simulationandresults} we present our model-based simulation results.
In section~\ref{section:relatedwork}, we describe related work in this subject to place our contributions in context. We conclude in section~\ref{section:conclusionandfuturework} and discuss future work.

\section{Backpressure with Adaptive Redundancy}
\label{section:BWAR}

In this section, we first describe traditional backpressure scheduling and routing and then our new proposal for backpressure scheduling/routing with adaptive redundancy (BWAR).
In both cases, we assume that there are $N$ nodes in the network, and time is discretized. We assume a multi-commodity flow system in which every node could be a potential destination (corresponding to a particular commodity).

\subsection{Traditional Backpressure Scheduling and Routing}
\label{TraditionalBackpressureSchedulingandRouting}
We assume that each node maintains $N-1$ queues, one for each commodity, with the $j^{th}$ queue at each node containing packets that are destined for node $j$. Let $Q_i^c(t)$ indicate the number of packets destined to node $c$ queued at node $i$ at time $t$. Naturally, $Q_i^i(t)=0 ~\forall t$. Let $\mu_{ij}^c(t)$ be the scheduling and routing variable that indicates the number of packets of commodity $c$ to be scheduled on link $(i,j)$. Traditional backpressure scheduling/routing~\cite{182479, Georgiadis:2006:RAC:1166401.1166402} selects the $\mu_{ij}^c(t)$ that solve the following problem (a form of maximum weight independent set selection):

\begin{align}
\max \sum\limits_{i,j,c} \Delta_{ij}^c (t) \cdot \mu_{ij}^c (t) \nonumber
\end{align}
\vspace{-0.6cm}
\begin{align}
\label{backpressureMax}
&\subjectto, \nonumber \\
&\sum_c\mu_{ij}^c (t) \le \theta_{ij}(t), & \forall i, \forall j \nonumber \\
&\mu_{ij}^c(t) \cdot \mu_{km}^d(t) = 0, &\left((i,j),(k,m)\right) \in \Omega(t), \forall c, \forall d
\end{align}

Where $\Delta_{ij}^c (t) = Q_i^c (t) - Q_j^c (t)$ is the link weight, which denotes the queue differential for commodity $c$ on link $(i,j)$ at slot $t$ and the feasibility constraints on $\mu_{ij}^c (t)$ pertain to the available network capacity, taking into account the interference between nodes. $\theta_{ij}(t)$ is the channel state in terms of number of packets that can be transmitted over link $(i,j)$ during slot $t$. $\Omega(t)$ is the link interference set at slot $t$ such that if link $(i,j)$ interferes with link $(i',j')$ at slot $t$ then  $\left((i,j), (i',j')\right) \in \Omega(t)$ and hence, those two links can not be both scheduled at slot $t$. The maximization problem in~\eqref{backpressureMax} can be solved by finding the maximum commodity $c^*_{ij}(t)$ for each link $(i,j)$ at slot $t$ that maximizes $\Delta_{ij}^c (t)$ and assign $\mu_{ij}^c (t) = 0$ for all $c \ne c^*_{ij}(t)$ and then solve,

\begin{align}
\max \sum\limits_{i,j} \Delta_{ij}^{c^*_{ij}(t)} (t) \cdot \mu_{ij}^{c^*_{ij}(t)} (t) \nonumber
\end{align}
\vspace{-0.6cm
}
\begin{align}
\label{backpressureMaxmaxcom}
&\subjectto, \nonumber \\
&\mu_{ij}^{c^*_{ij}(t)} (t) \le \theta_{ij}(t), & \forall i, \forall j \nonumber \\
&\mu_{ij}^{c^*_{ij}(t)}(t) \cdot \mu_{km}^{c^*_{km}(t)}(t) = 0, &\left((i,j),(k,m)\right) \in \Omega(t)
\end{align}

\subsection{BWAR Scheduling and Routing}

Our proposed enhancement of backpressure with adaptive redundancy works as follows. We have an additional set of $N-1$ duplicate buffers of size $D_{max}$ at each node. Besides the original queue occupancy $Q_i^c(t)$ which has the same meaning as in traditional backpressure, the duplicate queue occupancy is denoted by $D_i^c(t)$, that indicates the number of duplicate packets at node $i$ that are destined to node $c$ at time $t$. Again, $Q_i^i(t)= D_i^i(t) = 0 ~\forall t$ since destinations need not buffer any packets intended for themselves.  The duplicate queues are maintained and utilized as follows:

\begin{itemize}
\item Original packets when transmitted are removed from the main queue; however, if the queue size is lower than a certain threshold $q_{th}$, then the transmitted packet is duplicated and kept in the duplicate buffer associated with its destination if it is not full otherwise no duplicate is created. We found that setting both $q_{th}$ and $D_{max}$ to the value of the maximum link service rate is enough and gives superior delay results.
\item Duplicate packets are not removed from the duplicate buffer when transmitted. They are only removed when they are notified to be received by the destination, or a pre-defined timeout has occurred.
\item When a certain link is scheduled for transmission, the original packets in the main queue are transmitted first. If no more original packets are left, only then duplicates are transmitted. Thus the duplicate queue has a strictly lower priority.
\end{itemize}

Similar to original backpressure scheduling/routing, the BWAR scheduling/routing also requires the solution of a similar maximum weight independent set problem:

\begin{align}
\max \sum\limits_{i,j,c} \Delta_{\text{BWAR},ij}^c (t) \cdot \mu_{ij}^c(t) \nonumber
\end{align}
\vspace{-0.6cm}
\begin{align}
\label{BWARMax}
&\subjectto, \nonumber \\
&\sum_c\mu_{ij}^c (t) \le \theta_{ij}(t),& \forall i, \forall j \nonumber \\
&\mu_{ij}^c(t) \cdot \mu_{km}^d(t) = 0,&\left((i,j),(k,m)\right) \in \Omega(t), \forall c, \forall d
\end{align}

We define an enhanced link weight for BWAR, $\Delta_{\text{BWAR},ij}^c (t)$ as follows, to take into account the occupancy of the duplicate buffer.

\begin{align}
\label{BWARdelta}
\Delta_{\text{BWAR},ij}^c (t) = \left(Q_i^c (t) - Q_j^c (t)\right) + \frac{1}{2}  \left(\mathbf{1}_{j=c \text{~And~} Q_i^c(t)+D_i^c(t)>0}\right) \nonumber
\\+ \frac{1}{4} \frac{1}{D_{max}} \left(D_i^c (t) - D_j^c (t)\right)
\end{align}
Here the indicator function $\mathbf{1}_{j=c \text{~And~} Q_i^c(t)+D_i^c(t)>0}$ denotes that node $j$ is the final destination for the considered commodity $c$. This gives higher weight to commodities that encounter their destinations. We show later how this effectively results in dramatic delay improvement. Similarly, the maximization problem in~\eqref{BWARMax} can be solved first by finding the maximum commodity $c^*_{\text{BWAR},ij}(t)$ for each link $(i,j)$ at slot $t$ that maximizes $\Delta_{\text{BWAR},ij}^c (t)$ followed by the same approach discussed earlier in \ref{TraditionalBackpressureSchedulingandRouting}. It is important to notice that a solution to \eqref{BWARMax} is indeed a solution to \eqref{backpressureMax} assuming that $Q_i^c(t)$ and $\mu_{ij}^c(t)$ are integers. The small weight added in \eqref{BWARdelta} gives advantage first to links/commodities which encounter the destination and then to higher duplicate buffer deferential to increase the chance of serving duplicates. The small fractions in~\eqref{BWARdelta} assures this priority when there are ties in~\eqref{backpressureMax} to boost delay performance.

\subsection{Backpressure routing in intermittently connected networks}

In general backpressure scheduling is NP-hard, owing to the MWIS problem that needs to be solved at each time. However, in this paper, we focus on intermittently connected networks, that consist of sparse encounters between pairs of nodes. Therefore, at any given time, the size of any connected component of the network is very small. In this case, the scheduling problem is dramatically simplified.

\subsection{Practical Duplicate Removal}

As can be seen from the above description, BWAR creates duplicate packets whenever the transmitter's queue occupancy is low. In an ideal setting, for efficiency, the duplicated packets in the network should be deleted instantaneously when any copy is delivered to the intended destination. This could only be implemented practically in intermittently connected networks where a centralized control plane is available that can provide such an instantaneous acknowledgement to all nodes in the network. In other cases, some other mechanism is sought, so we propose the following timeout mechanism. Whenever a packet arrives into the network, it is time-stamped. After a timeout period $P$ from that arrival time, any duplicate copies of that packet at any node in the network will be deleted. To obtain higher delay performance improvement, when an original packet is duplicated, it is placed in the duplicated buffer giving it lower service priority, however, it is flagged and not deleted when a timeout occurred. It is only removed when it gets acknowledged directly by the destination.

In the next section we undertake an analysis of the performance of BWAR and compare it with the known results for traditional backpressure routing. Specifically, we prove that any feasible rate vector is also stabilized by BWAR, and the bound that we can give on the expected queue occupancy for BWAR is better than that for regular backpressure.

\section{Mathematical Analysis}

\subsection{Review of the Analysis of Basic Backpressure}
\label{section:reviewofbackpressure}
We consider a timeslotted network with N nodes that
communicate with each other.  Packets arrive to each
node, and each packet must be delivered to a specific
destination, possibly via a multi-hop path.  Each node
maintains \emph{several} queues, one per destination, to store
packets. Each queue has the following dynamics:
\begin{align}
\label{QuD1_1}
Q(t+1) = \max[Q(t)-\mu(t),0] + A(t)
\end{align}

Where $Q(t)$ is the queue
size at time $t$,
$\mu(t)$ is the transmission rate
out of the queue at time $t$, and
$A(t)$ is the total packet arrivals to the queue at
time $t$.

Each time slot, we observe the queue states and the channel
states and make scheduling and routing decisions
based on this information. To clear this out, let $Q_n^c(t)$ be the queue backlog (number
of packets) in node $n \in \{1, 2, ..., N\}$ that are destined for node $c \in \{1, ..., N\} \backslash \{n\}$
at slot $t$. Let $A_n^c(t)$ be the exogenous packet arrivals that come to node $n$ and
destined to node $c$ at time
$t$ with rate $\lambda_n^c$. Exogenous arrivals are the packets that just entered the network. Endogenous arrivals, however, are arrivals from other nodes and were already inside the network.
Packets may be forwarded to several nodes before reaching the destination.
Let us define the capacity region $\Lambda$ to be the set of all possible arrival rate vectors $(\lambda_n^c)_{n,c}$ that are stabilizing by some scheduling and routing strategy.
Let $\theta_{ab}(t)$ be the channel state from node $a$ to node $b$ at time $t$ in terms of how many packets can be transmitted.
Let $\mu_{ab}(t)$  be the scheduled service rate from node $a$ to node $b$ at slot $t$.
Let $\mu_{ab}^c(t)$ be the service rate for commodity $c$ routed from node $a$ to node
$b$ at time
$t$ and must satisfy:
\begin{align}
\label{QuD1_2}
\sum_c{\mu_{ab}^c(t)} \le \mu_{ab}(t) \le \theta_{ab}(t)
\end{align}

The queue dynamics for each time slot and for each queue is the
following:
\begin{align}
\label{QuD2_1}
Q_n^c(t+1) = \max[Q_n^c(t)-\sum_b{\mu_{nb}^c(t)},0]
\nonumber \\ + A_n^c(t) + \sum_a{\tilde{\mu}_{an}^c(t)}
\end{align}

Where $\tilde{\mu}$ is the actual transfer rate due to insufficient packets in the
queue. For example, on some slots we may able to send 5
packets, but we only send 3, because only 3 were available in the queue. In equation~\eqref{QuD2_1}, $A_n^c(t)$ are the exogenous arrivals and $\sum_a{\tilde{\mu}_{an}^c(t)}$ are the endogenous
arrivals to node $n$.

Define the vector $\textbf{Q}(t) = (Q_n^c(t))_{n,c}$ to be the
vector of all queues in the network at time $t$. The Lyapunov function
$L(\textbf{Q}(t))$ can be defined as following:
\begin{align}
L(\mathbf{Q}(t)) = \sum_{n,c} Q_n^c(t)^2
\end{align}

The Lyapunov drift $\Delta(\mathbf{Q}(t))$ is defined as following:
\begin{align}
\Delta(\mathbf{Q}(t)) = \mathbb{E} \{L(\mathbf{Q}(t+1))-L(\mathbf{Q}(t))|\mathbf{Q}(t)\}
\end{align}

It has been already proven by~\cite{Georgiadis:2006:RAC:1166401.1166402, 182479} that:

\begin{align}
\label{Drift_1}
\Delta(\mathbf{Q}(t))\le
\sum_{n,c} \mathbb{E}\left\{\beta_n^c(t) \right\} - 2 \sum_{n,c}{ Q_n^c(t) \mathbb{E} \left \{ \psi_n^c(t)\middle \vert \mathbf{Q}(t) \right \}}
\end{align}

Such that:
\begin{align}
\label{betaeq}
\beta_n^c(t)=\left(\sum_b \mu_{nb}^c(t)\right)^2 + \left(A_n^c(t) + \sum_a \mu_{an}^c(t)\right)^2
\end{align}
and,
\begin{align}
\label{psieq}
\psi_n^c(t) = \sum_b{\mu_{nb}^c(t)} -\sum_a{\mu_{an}^c(t)} -A_n^c(t)
\end{align}
Maximizing $\sum_{n,c}{ Q_n^c(t) \mathbb{E} \left \{ \psi_n^c(t)\middle \vert \mathbf{Q}(t) \right \}}$  in \eqref{Drift_1} which is equivalent to the maximization problem defined in~\eqref{backpressureMax}
yields the backpressure algorithm for scheduling and routing and it
has been proven by~\cite{Georgiadis:2006:RAC:1166401.1166402, 182479} that it supports the maximum capacity $\Lambda$.
The average queue occupancy bound for backpressure scheduling and routing is:
\begin{align}
\label{RBbound}
\bar{Q}  \le \frac{\bar{\beta}}{2\epsilon}
\end{align}
such that,
\begin{align}
\label{Qbareq}
\bar{Q} &=& \lim_{T \to \infty} \frac{1}{T}\sum_{\tau=0}^T \sum_{n,c}\mathbb{E}\left\{Q_n^c(\tau)\right\} \\
\label{betabareq}
\bar{\beta}  &=& \lim_{T \to \infty} \frac{1}{T} \sum_{\tau=0}^T \sum_{n,c}\mathbb{E}\left\{\beta_n^c(\tau)\right\} \\
\label{epsiloneq}
\epsilon &=& \argmax_{x \ge 0} (\lambda_n^c+x)_{n,c} \in \Lambda
\end{align}

Where, $\bar{Q}$ is the average of total queue backlog occupancy. $\bar{\beta}$ is the sum of the second moment of the scheduled transmission rate out of each queue plus the second moment of the sum of the arrivals and scheduled transmission rate into each queue and summed over all queues. $\epsilon$ is the maximum positive number such that adding $\epsilon$ to each arrival rate still makes them inside the capacity region $\Lambda$.

\subsection{Analysis of BWAR}
\label{section:bwarqueueingdynamics}
Here is a formal mathematical description of backpressure
with adaptive redundancy. As before, let $Q_n^c(t)$ to be queue backlog in node
$n$ of commodity $c$ at time
slot $t$. We define $D_n^c(t)$ to be number of redundant packets in node
$n$ of commodity $c$ at time
$t$. Redundant packets are stored separately in
redundant buffers. Redundant packets have lower priority in such a way
that no redundant packet is served unless the queue of original packets is
empty.
For all time slots $t$, $A_n^c(t)$, $\theta_{ab}(t)$, $\mu_{ab}(t)$, $\mu_{ab}^c(t)$ and $\tilde{\mu}_{ab}^c(t)$ are defined
exactly as before. Arrival rates $\lambda_n^c$ are also defined as before.
The queue
dynamics in equation \eqref{QuD2_1} is updated for adaptive redundancy to be:

\begin{align}
\label{QDBWARE}
Q_n^c(t+1) = \max [Q_n^c(t) - \gamma_n^c(t) - \sum_b \mu_{nb}^c(t) , 0]
\nonumber \\ + A_n^c(t) + \sum_a \tilde{\mu}_{an}^c(t)
\end{align}

Where $\gamma_n^c(t)$ is the number of original packets inside node $n$ of commodity $c$ at time slot $t$ that are known to be delivered by some duplicates to the destination using our BWAR strategy.
One ideal model is that we find out which packets are delivered immediately, another is that we
find out after some delay.  Our analysis allows for any such knowledge of delivered packets. We show later a practical timeout-based strategy for duplicate removals.
Those $\gamma_n^c(t)$ packets are needed to be removed from the queue since they are already known to be delivered.
We assume that the deletion happens during the time slot $t$ hence at the beginning of time slot $t$ none of those packets are deleted yet but are known to be deleted.
The queue dynamics in \eqref{QDBWARE} consider only original packets and does not take into account the duplicate packets.
We define the redundant buffer dynamics that are isolated from the original queue dynamics as following:

\begin{align}
\label{BufD1_1}
 D_n^c(t+1) = D_n^c(t)- \tilde{\gamma}_n^c(t) + \delta_n^c(t) +  \sum_a{\omega_{an}^c(t)}
\end{align}

Where $\tilde{\gamma}_n^c(t)$ denotes the number of duplicates in node $n$
of commodity $c$ at time $t$
that are known to be already delivered to the destination and hence they must be removed.
$\delta_n^c(t)$ is number of duplicates created at node $n$ during slot $t$ according to the adaptive redundancy criteria.
$\omega_{ab}^c(t)$ is the actual duplicate transmissions from node
$a$ to node $b$ of commodity
$c$ at time $t$. BWAR algorithm chooses $\delta_n^c(t)$ and $\omega_{ab}^c(t)$  in such away to assure that $D_n^c(t+1) \le D_{max}$~~ $\forall t$.

As before, $\mathbf{Q}(t) = (Q_n^c(t))_{n,c}$ is the vector of all queue backlogs at time $t$.
Let $U_n^c(t)$ to be the undelivered
queue backlog in node $n$ of commodity $c$ at time $t$. Hence,
\begin{align}
 \label{undeliveredeq}
 U_n^c(t) = Q_n^c(t) - \gamma_n^c(t)
\end{align}

Let $\mathbf{U}(t) = (U_n^c(t))_{n,c}$ be the vector of all queue backlogs of undelivered packets at time $t$.
Let $\mathbf{\Gamma}(t) = (\gamma_n^c(t))_{n,c}$ be the vector of all removed duplicates at time $t$. Define the Lyapunov function $L(\mathbf{X}) = \sum (X_i)^2$. Assume that $\bar{Q}$, $\bar{\beta}$ and $\epsilon$ are defined as before in~\eqref{Qbareq}, \eqref{betabareq} and \eqref{epsiloneq} respectively.

Let also define,
\begin{align}
\bar{U} &=& \lim_{T \to \infty} \frac{1}{T}\sum_{\tau=0}^T \sum_{n,c}\mathbb{E}\left\{U_n^c(\tau)\right\} \\
\overline{\Gamma^2} &=& \lim_{T \to \infty} \frac{1}{T} \sum_{\tau=0}^T \sum_{n,c}\mathbb{E}\left\{\left(\gamma_n^c(\tau)\right)^2 \right\} \\
\overline{Q.\Gamma} &=& \lim_{T \to \infty} \frac{1}{T} \sum_{\tau=0}^T \sum_{n,c}\mathbb{E}\left\{Q_n^c(\tau).\gamma_n^c(\tau) \right\}\\
\overline{U.\Gamma} &=& \lim_{T \to \infty} \frac{1}{T} \sum_{\tau=0}^T \sum_{n,c}\mathbb{E}\left\{U_n^c(\tau).\gamma_n^c(\tau) \right\}
\end{align}

Where, $\bar{U}$ is the average of total queue backlog occupancy for undelivered packets in the main queues.
$\overline{\Gamma^2}$ is the second moment of number of removed packets in each original queue because those packets are known that are delivered by duplicates to the destination and summed over all queues.
$\overline{Q.\Gamma}$ is the joint second moment of number of removed packets and the queue backlog summed over all queues.
$\overline{U.\Gamma}$ is the joint second moment of number of removed packets and the queue backlog of undelivered packets summed over all queues.

For simplicity of exposition, we prove
the result in the simple case when arrival rates $A_n^c(t)$
and the channel states $\theta_{ab}(t)$ are i.i.d. over slots.  This
can be extended to general ergodic (possibly non-i.i.d.) processes
using a T-slot drift argument as in \cite{neely-power-network-jsac}.

\newtheorem{theorem}{Theorem}
\begin{theorem}
If the channel states $\theta_{ab}(t)$ are i.i.d. and the arrival processes $A_n^c(t)$ are i.i.d. with rates $\lambda_n^c$ that are inside the capacity region $\Lambda$ such that $(\lambda_n^c + \epsilon)_{n,c}\in \Lambda$ for some $\epsilon>0$, then BWAR stabilizes all queues with the following bound on the average of total queue occupancy of undelivered packets $\bar{U}$,
\begin{align}
\bar{U} \le \frac{\bar{\beta} - \overline{\Gamma^2} - 2\overline{U.\Gamma}}{2\epsilon}
\end{align}
\end{theorem}

\begin{IEEEproof}
Squaring both sides of \eqref{QDBWARE},
\begin{align}
Q_n^c(t+1)^2 \le  \left(Q_n^c(t) - \gamma_n^c(t)\right)^2 + \beta_n^c(t)
\nonumber \\
 - 2\left(Q_n^c(t) - \gamma_n^c(t)\right)\psi_n^c(t)
\end{align}

where $\beta_n^c(t)$ and $\psi_n^c(t)$ are defined as before in~\eqref{betaeq}~and~\eqref{psieq} respectively.

Summing over all $n$ and $c$,
\begin{align}
\sum_{n,c}Q_n^c(t+1)^2  \le  \sum_{n,c}\left(Q_n^c(t) - \gamma_n^c(t)\right)^2 +\sum_{n,c} \beta_n^c(t) \nonumber \\
 - 2\sum_{n,c}\left(Q_n^c(t) - \gamma_n^c(t)\right)\psi_n^c(t)
\end{align}

Taking the conditional expectation $\mathbb{E}\{.|\mathbf{Q}(t)-\mathbf{\Gamma}(t)\}$,
\begin{align}
&\mathbb{E}\left\{L(\mathbf{Q}(t+1)) - L(\mathbf{Q}(t)-\mathbf{\Gamma}(t)) \middle|\mathbf{Q}(t)-\mathbf{\Gamma}(t)\right\} \le
\nonumber \\
& \mathbb{E}\left\{ \sum_{n,c} \beta_n^c(t) - 2\sum_{n,c}\left(Q_n^c(t) - \gamma_n^c(t)\right)
\psi_n^c(t) \middle|\mathbf{Q}(t)-\mathbf{\Gamma}(t)\right\}
\end{align}

Since our BWAR policy maximizes~\eqref{BWARMax} and hence~\eqref{backpressureMax} taking into account the undelivered packets $\mathbf{U}(t)$ only, it will also maximize:
\begin{align}
\label{BWARmaxpolicy}
\mathbb{E}\left\{\sum_{n,c}\left(Q_n^c(t) - \gamma_n^c(t)\right)\psi_n^c(t) \middle|\mathbf{Q}(t)-\mathbf{\Gamma}(t)\right\}
\end{align}

However, because $(\lambda_n^c + \epsilon)_{n,c}$ are inside the capacity region $\Lambda$,
we know from \cite{neely-power-network-jsac} that there exists a stationary and randomized algorithm $alg^*$, which makes decisions
independent of $\mathbf{Q}(t) - \mathbf{\Gamma}(t)$, yielding ${\psi^*}_n^c(t)$ that satisfy:
\[ \mathbb{E}\left\{{{\psi^*}_n^c(t)}\right\} \leq -\epsilon \: \: \forall n, c \]
Because BWAR maximizes \eqref{BWARmaxpolicy}, it follows that:
\begin{align}
&\mathbb{E}\left\{\sum_{n,c}(Q_n^c(t) - \gamma_n^c(t))\psi_n^c(t)\middle|\mathbf{Q}(t)-\mathbf{\Gamma}(t)\right\} \leq \nonumber \\
&\mathbb{E}\left\{\sum_{n,c}(Q_n^c(t) - \gamma_n^c(t)){\psi^*}_n^c(t)\right\} = \nonumber \\
&-\sum_{n,c}(Q_n^c(t) - \gamma_n^c(t))\epsilon
\end{align}

Using this in (27) yields,
\begin{align}
\mathbb{E}\left\{L(\mathbf{Q}(t+1)) - L(\mathbf{Q}(t)-\mathbf{\Gamma}(t)) \middle|\mathbf{Q}(t)-\mathbf{\Gamma}(t)\right\} \le
\nonumber \\
  \sum_{n,c} \mathbb{E}\left\{\beta_n^c(t) \middle|\mathbf{Q}(t)-\mathbf{\Gamma}(t)\right\}
- 2\epsilon \sum_{n,c}\left(Q_n^c(t) - \gamma_n^c(t)\right)
\end{align}

Taking iterative expectation,
\begin{align}
\mathbb{E}\left\{L(\mathbf{Q}(t+1)) \right\}
 - \mathbb{E}\left\{L(\mathbf{Q}(t)-\mathbf{\Gamma}(t)) \right\} \le
\nonumber \\
   \sum_{n,c} \mathbb{E}\left\{\beta_n^c(t)\right\}
- 2\epsilon \sum_{n,c}\mathbb{E}\left\{\left(Q_n^c(t) - \gamma_n^c(t)\right)\right\}
\end{align}

Notice that:
\begin{align}
\mathbb{E}\left\{L(\mathbf{Q}(t)-\mathbf{\Gamma}(t)) \right\} =
\mathbb{E}\left\{L(\mathbf{Q}(t)\right\}
+ \mathbb{E}\left\{L(\mathbf{\Gamma}(t)) \right\}
\nonumber \\
- 2 \mathbb{E}\left\{\mathbf{Q}(t).\mathbf{\Gamma}(t) \right\}
\end{align}

Hence by summing over time slots $\tau \in \{0, ..., T\}$ and by telescoping,
\vspace{-0.4cm}
\begin{align}
&\mathbb{E}\left\{L(\mathbf{Q}(T)) \right\}
 - \mathbb{E}\left\{L(\mathbf{Q}(0))\right\}
 -\sum_{\tau=0}^T \mathbb{E}\left\{L(\mathbf{\Gamma}(\tau)) \right\}
\nonumber \\
& + 2 \sum_{\tau=0}^T \mathbb{E}\left\{ \mathbf{Q}(\tau).\mathbf{\Gamma}(\tau) \right\}
\le
\nonumber \\
& \sum_{\tau=0}^T  \sum_{n,c} \mathbb{E}\left\{\beta_n^c(\tau)\right\}
- 2\epsilon \sum_{\tau=0}^T \sum_{n,c}\mathbb{E}\left\{\left(Q_n^c(\tau) - \gamma_n^c(\tau)\right)\right\}
\end{align}

Dividing by $T$ and taking the $\lim$ for $T \to \infty$ implies:

\begin{align}
\label{mainresultQ}
\bar{Q} - \bar{\Gamma} \le \frac{\bar{\beta} + \overline{\Gamma^2} - 2\overline{Q.\Gamma}}{2\epsilon}
\end{align}

Now for undelivered packets $\bar{U}$, we have by~\eqref{undeliveredeq} and~\eqref{mainresultQ},

\begin{align}
\nonumber
\bar{U} \le \frac{\bar{\beta} - \overline{\Gamma^2} - 2\overline{U.\Gamma}}{2\epsilon}
\end{align}
\end{IEEEproof}

\textbf{Remark:} Note that the computation of $\overline{\Gamma^2}$ and $\overline{U.\Gamma}$ is determined by the duplicate removal strategies. Depending on these terms, the queue bound in this above theorem could be much lower than the queue occupancy bound for regular backpressure in~\eqref{RBbound}. Thus we have a formal guarantee that BWAR is no worse in terms of throughput than backpressure, and potentially much better in terms of delay, since by Little's theorem average delay is proportional to the average number of undelivered packets. We will validate this finding with model in the next section.

\section{Model-based Simulations}
\label{section:simulationandresults}

\subsection{The Cell-Partitioned Model}

The model in this paper simplifies the control variables to be the whole transmission rates $\mu_{ab}(t)$ for
scheduling and the
commodity transmission rates $\mu_{ab}^c(t)$ for routing.

We simulate BWAR in the context of encounter-based scheduling and routing for a simple model (cell-partitioned network), which yields
useful insights on its performance.  In this idealized model the network deployment area is separated
into disjoint cells and nodes have i.i.d. mobility model~\cite{1435642} as follows.
We have $N$ nodes and $C$
cells. At each slot $t$, node
$n$ can be inside any cell with equal probabilities of
$\frac{1}{C}$. For collision and interference simplicity, only one transmission (one packet) is allowed in each
cell in each time slot. Because of this we set $q_{th} = D_{max} = 1$. Another simplifying assumption is that the nodes in the network are organized into pairs,
acting as destinations to each other. Each node has Bernoulli exogenous arrivals intended for its pair. Depending on the number
of cells $C$ in
the network we can choose the right number of the nodes $N \approx 1.79 \cdot C$ in order to maximize throughput as shown in~\cite{1435642}. Our simulation results show that by
optimizing number of nodes based on the number of cells to maximize throughput, the delay also is improved. We consider in our simulations, networks of sizes 9, 12, 16, 20, and 25 cells in the network. And for
optimality, number of nodes are chosen to be 16, 20, 28, 34, and 44
respectively. For timeout duplicate removals we set the timeout value $P = C$.

Here we show how BWAR works in the
cell-partitioned network with the simplifying assumption that only one transmission is allowed
per cell per time slot. Each time slot $t$ and for
each cell $l$ we choose two nodes
$a^*$ and $b^*$ and commodity $c^*$ such
that:

\begin{itemize}
	\item $a^*$ and $b^*$ are in cell
$l$.
	\item $Q_{a^*}^{c^*}(t) - Q_{b^*}^{c^*}(t) \ge Q_a^c(t) - Q_b^c(t)$; for all $c$, for
all $a$ and $b$ in cell
$l$ at time slot $t$. This captures the maximization of queue differentials of the main queues.
	\item If there exists $a, b$ in cell
$l$ such that, \\ $Q_a^b(t) - Q_b^b(t) = Q_{a^*}^{c^*}(t) - Q_{b^*}^{c^*}(t)$ then $c^* = b^*$. This captures the destination advantage.
	\item If there exists $a, b$ in cell
$l$ and $c$ such that \\ $Q_a^c(t) - Q_b^c(t) = Q_{a^*}^{c^*}(t) - Q_{b^*}^{c^*}(t) $  and \\ $\{ c^* \ne b^*$ or $[(c=b)$ and $(c^*=b^*)] \} $   then \\
$(Q_a^c(t)+D_a^c(t))-(Q_b^c(t)+D_b^c(t)) \le (Q_{a^*}^{c^*}(t)+D_{a^*}^{c^*}(t))-(Q_{b^*}^{c^*}(t)+D_{b^*}^{c^*}(t))$. This captures the maximization of duplicate buffer differentials if there are some ties in main queue differentials.
\end{itemize}

The algorithm simply assigns $\mu_{a^*b^*}^{c^*}(t)$ a value of 1, and assigns all
other $\mu_{ab}^c(t)$ a value of 0 such that $a,b$ in cell
$l$.

When a transmission is made from node $a$ to node $b$ of commodity $c$ at time slot $t$ and that transmission will make $Q_a^c(t+1) + D_a^c(t+1) = 0$ then this transmitted packet is duplicated and stored in the duplicate buffer of node $a$ making $D_n^c(t) = 1$ instead of $0$. Duplicate packets are served only if there are no original packets to transmit. There is strict lower priority of duplicate packets compared to original packets.

\subsection{Protocol Variants}

\begin{figure*}[t]
\centering
\subfigure[Delay as we vary $N$ for low $\lambda = 0.001$]
{\includegraphics[width=0.4\textwidth]{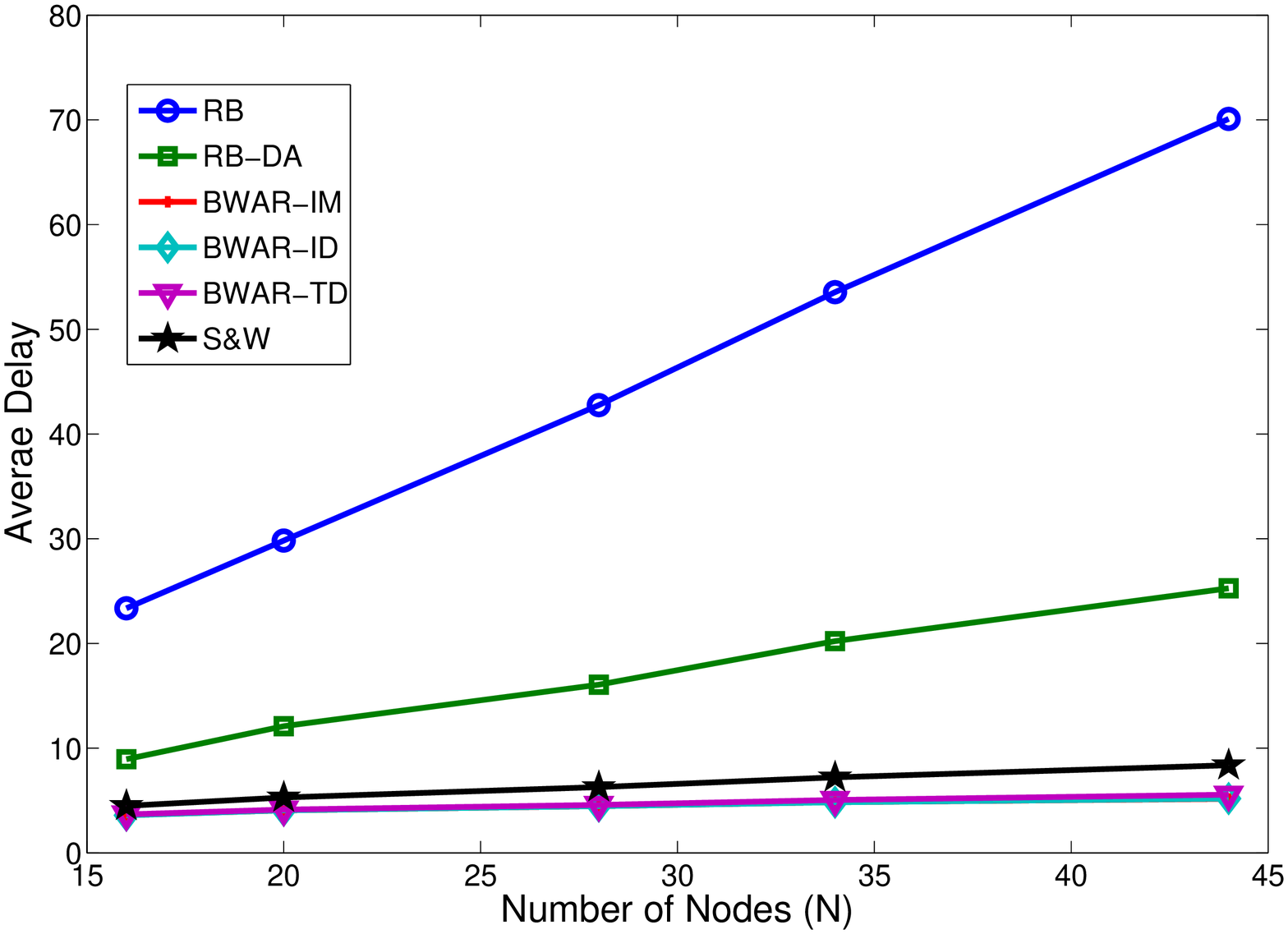}\label{fig:cellpartitioneddelay}}
\subfigure[Delay as we vary $\lambda$ for $N = 44$ of backpressure variants]
{\includegraphics[width=0.4\textwidth]{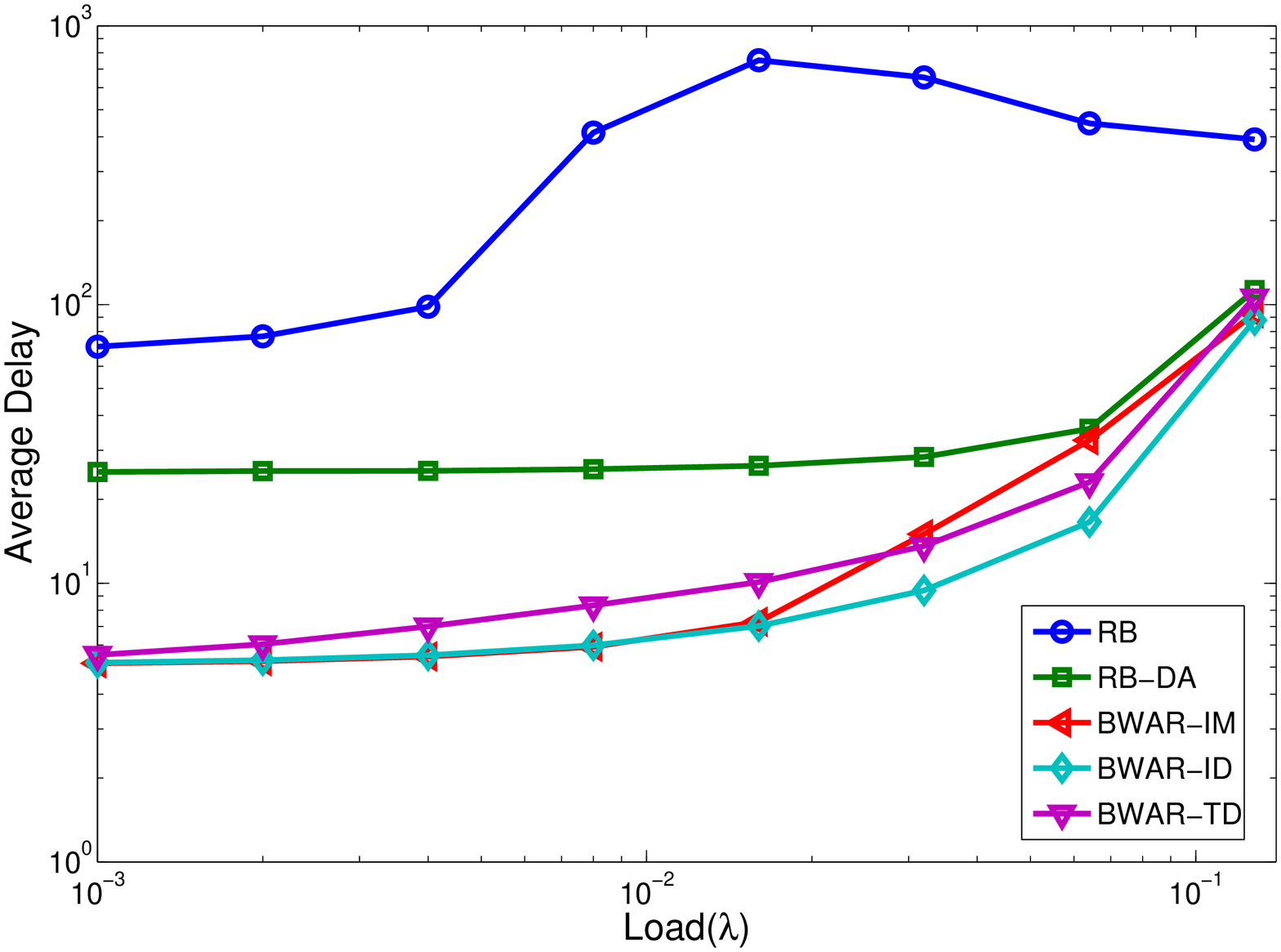}\label{fig:cellpartitioneddelay_load}}
\caption{Comparing delay performance of protocol variants: RB, RB-DA, BWAR-IM, BWAR-ID, BWAR-TD and S\&W under the cell-partitioned model.}
\label{fig:celldelay}
\end{figure*}

In the simulations, we implement and compare five different routing protocol variants. They are described as follows:

\begin{itemize}
\item \textbf{Regular Backpressure (RB)}: This is the basic backpressure scheduling and routing mechanism, where decisions are made purely based on queue differentials.
\item \textbf{Regular Backpressure with Destination Advantage (RB-DA)}: This is a slight modification in which packets corresponding to the destination are prioritized when the destination is encountered. As we show, this already yields significant delay improvements over regular backpressure.
\item \textbf{BWAR with Ideal packet removal and original packets retained in the Main queue (BWAR-IM)}: This is our novel backpressure with adaptive redundancy in which the destination advantage is also holds. Here, when an original packet is duplicated the original packet remains in the main queue while the duplicate is stored in the duplicate buffer. We assume here whenever a packet reaches the destination, all of its duplicates are deleted including the original one in the main queue instantaneously.
\item \textbf{BWAR with Ideal packet removal and original packets moved to Duplicate buffer upon copy (BWAR-ID)}: This is very similar to BWAR-IM. The only difference is that whenever an original packet is duplicated both the original packet and the duplicate are stored in the duplicate buffer (of course in two different nodes one in the receiver and the other in the sender respectively).
\item \textbf{BWAR with Time-out based packet removal and original packets moved to Duplicate buffer upon copy (BWAR-TD)}: This is a practical implementation of BWAR in which duplicates are deleted from the duplicate buffer after a predefined timeout value $P$ has passed since the first time the original packet is admitted to the network. However, the original packet that is kept in duplicate buffer is flagged and will not be deleted when a timeout occurred. It is only deleted if it gets acknowledged directly by the destination if its already received or otherwise it moved back to the main queue when it encounters the destination.
\item \textbf{Spray and Wait (S\&W)}: This is not a backpressure based mechanism. Spray and Wait is presented by T. Spyropoulos \emph{et al.}~\cite{Spyropoulos:2008:ERI:1373452.1373459} which is a state of the art routing scheme in intermittently connected mobile networks. S\&W creates a predefined fixed number of copies (spraying) of the packet when admitted to the network. Those copies are distributed to distinct nodes and then each copy waits until it encounters the destination. We implemented S\&W for comparison with BWAR. Our results show that BWAR outperforms S\&W especially in high load scenarios.
\end{itemize}

The evaluations are conducted using a custom simulator written in C++ (for repeatability, we make our code available online at http://anrg.usc.edu/downloads/). Each simulation runs for one million time slots.

In figure~\ref{fig:cellpartitioneddelay}, we show average delay of all above protocol variants as number of nodes $N$ vary for low load $\lambda = 0.001$ out of the per node capacity region $\Lambda_{\text{node}} = [0, 0.14]$. Delay is reduced significantly when BWAR is used. For this low load scenario all BWAR variants have almost the same average delay and they perform slightly better than Spray and Wait. Figure~\ref{fig:cellpartitioneddelay} also shows the great dramatic delay improvement of destination advantage without any redundancy in RB-DA compared to regular backpressure RB.

Figure~\ref{fig:cellpartitioneddelay_load} compares the average delay of all variants of backpressure-based protocols as we vary the load.
As expected, as the load increases the delay improvement of BWAR declines compared to RB-DA. Figure~\ref{fig:cellpartitioneddelay_load} also shows how BWAR-ID performs much better compared to BWAR-IM beyond some threshold of load($\lambda$). This shows how moving the duplicated original packet to the duplicate buffer has great delay enhancement for high load scenarios.

\begin{figure}[htbp]
\centering
{\includegraphics[width=0.45\textwidth]{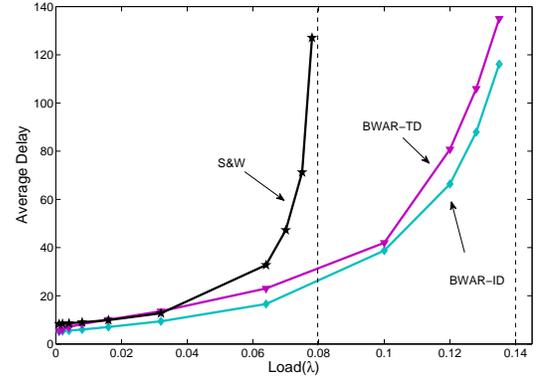}
\caption{Comparing S\&W delay with BWAR-ID and BWAR-TD as we vary $\lambda$ for $N = 44$}
\label{fig:cellpartitionedSprayAndFocus}}
\end{figure}

In Figure~\ref{fig:cellpartitionedSprayAndFocus}, results show how BWAR mechanism outperforms Spray and Wait (S\&W) delay performance for high load. It shows also how BWAR supports almost twice the capacity region of S\&W.

\begin{figure}[htbp]
	\centering
\includegraphics[width=0.45\textwidth]{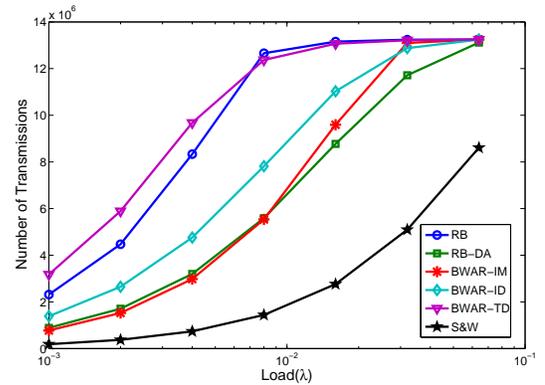}
	\caption{Comparing energy consumption as we vary $\lambda$ for $N = 44$ under the cell-partitioned model.}
\label{fig:cellpartitionedenergy_load}
\end{figure}

Surprisingly in figure
\ref{fig:cellpartitionedenergy_load}, BWAR-IM has a better total number of transmissions compared to regular backpressure RB-DA for low load despite
the flooding duplicates nature of BWAR at low load. Spray and Wait has superior energy consumption performance compared to all backpressure-based protocol variants considered. For future work, we intend to study the possibility of having both power optimization and adaptive redundancy features to be enabled on backpressure.

\begin{figure}[htbp]
	\centering
\includegraphics[angle=0, width=0.45\textwidth]{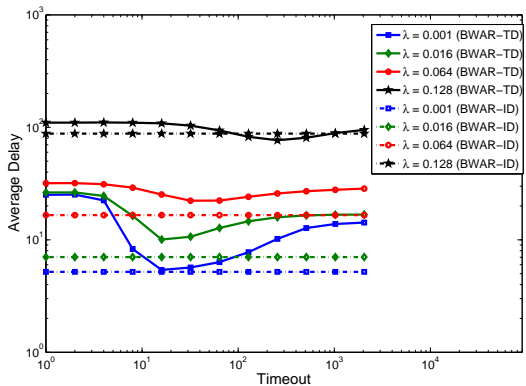}
\caption{Timeout effect of BWAR-TD and comparing it with BWAR-ID for different $\lambda \in \{0.001, 0.016, 0.064, 0.128\}$ for $N=44$ under the cell-partitioned model.}
\label{fig:cellpartitioneddelaytimeouteffect2}
\end{figure}

Figure~\ref{fig:cellpartitioneddelaytimeouteffect2} studies the effect of timeout value $P$ of BWAR-TD for removing duplicates under different load scenarios and compares its delay performance with ideal duplicate removals in BWAR-ID.

\section{Related Work}
\label{section:relatedwork}

The first theoretical work on backpressure scheduling is the classic result by Tassiulas and Ephremides in 1992, proving that this queue-differential based scheduling mechanism is throughput optimal (i.e., it can stabilize any feasible rate vector in a network)~\cite{182479}. Since then, researchers have combined the basic backpressure mechanism with utility optimization to provide a comprehensive approach to stochastic network optimization~\cite{Georgiadis:2006:RAC:1166401.1166402, 1665000, doi:10.2200/S00271ED1V01Y201006CNT007}.

Of most relevance to this work are papers on delay enhancements to backpressure. A number of papers~\cite{5549856, 1665003, 4796268} address the utility-delay tradeoff in optimization-oriented backpressure, to obtain a tradeoff based on a $V$ parameter such that the utility is improved by a factor of $O(1/V)$ while the delay is made to be polylogarithmic in $V$. Such a tradeoff has been shown to be practically achievable using LIFO queueing in~\cite{5930067}, at the cost of a small probability of dropping packets. The first-ever implementation of dynamic backpressure routing aimed for wireless sensor networks (BCP)~\cite{BCP} uses such a LIFO mechanism. As our focus in this work is not on utility optimization, the techniques presented in these works are somewhat orthogonal to the redundancy approach we develop here. Another set of papers~\cite{5062086, Neely2009862, Javidi10} consider the use of shortest path routing in conjunction with backpressure to improve the delay performance. These techniques are well suited for static networks in which such paths can be computed; however, since our focus is on encounter based networks with limited connectivity, such an approach is not applicable.

In~\cite{Bui09Infocom}, the authors present a mechanism whereby only one real queue is maintained for each neighbor, along with virtual counters/shadow queues for all destinations, and show that this yields delay improvements.  And in \cite{5935084}, a novel variant of backpressure scheduling mechanism is proposed which uses head of line packet delay instead of queue lengths as the basis of the backpressure weight calculation for each link/commodity, also yielding enhanced delay performance. However, these works both assume the existence of static fixed routes. It would be interesting to explore in future work whether their techniques can be applied to intermittently connected encounter-based mobile networks, and if so, how these approach can be further enhanced by the use of the adaptive redundancy that we propose in this work.

Ryu \emph{et al.} present two works on backpressure routing aimed specifically for cluster-based intermittently connected networks~\cite{Ryu_back-pressurerouting, Ryu:2010:BRR:1859995.1860037}. In \cite{Ryu_back-pressurerouting}, the authors develop a two-phase routing scheme, combining backpressure routing with source routing for cluster-based networks, separating intra-cluster routing from inter-cluster routing. They show that this approach results in large queues at only a subset of the nodes, yielding smaller delays than conventional backpressure. In~\cite{Ryu:2010:BRR:1859995.1860037}, the authors implement the above-mentioned algorithm in a real experimental network and show the delay improvements empirically. The key difference of these works from ours is that we do not make any assumption about the intermittently connected network being organized in a cluster-based hierarchy.

Dvir and Vasilakos~\cite{dvir:backpressure-based} also consider backpressure routing for intermittently connected networks, with link weights similar to that used in BCP~\cite{BCP}. They evaluate Weighted Fair Queueing in addition to LIFO and show through simulations that it offers energy improvements. Their work does not explicitly address additional delay improvements needed for these kinds of networks.

There is a rich literature on routing in delay tolerant / intermittently connected encounter based mobile networks (see~\cite{SpyropoulosSurvey} for a comprehensive survey). Although there exist single-copy routing mechanisms for such networks~\cite{4430783}, it has been well-recognized that replication is helpful in reducing delay. While basic epidemic routing~\cite{Vahdat00epidemicrouting} creates multiple message replicas for reliable, fast delivery, it incurs too high of a transmission cost. Smarter multi-copy routing mechanisms have therefore been developed such as Spray and Wait ~\cite{Spyropoulos:2008:ERI:1373452.1373459}, and SARP~\cite{SARP}. These works introduce redundant packet transmissions to improve delay. However, all of these approaches are not adaptive to the traffic and therefore will hurt the throughput performance of the network. This has been noted before, by the authors of \cite{Ryu:2010:BRR:1859995.1860037}, who write that ``replication-based algorithms such as epidemic routing for DTNs ... result in lower throughput since multiple copies of a piece of data need to be forwarded and stored (and therefore not throughput optimal).'' In fact, in~\cite{1435642}, it has been theoretically proved that capacity of such schemes that use fixed redundancy is necessarily lower. In this work, we present the first backpressure algorithm that uses replication in an adaptive manner so as to maintain throughput optimality while reducing delay. We explicitly compare our BWAR scheme with Spray and Wait, and show through our evaluation that not only does it provide similar, even better, delay performance, it does so without hurting throughput optimality; specifically, we show that BWAR can handle much higher traffic load than Spray and Wait.

To summarize, this paper on BWAR is the first work that explicitly combines the best of both worlds: multi-copy routing for intermittently connected networks and throughput-optimal backpressure scheduling. This combination yields better delay performance than traditional backpressure, particularly at low loads, and better ability to handle high traffic than traditional DTN/ICN routing schemes.

\section{Conclusion and Future Work}
\label{section:conclusionandfuturework}

We have presented in this paper BWAR, an enhanced backpressure algorithm that introduces adaptive
redundancy to improve delay performance. We have proved analytically that this algorithm
is also throughput optimal while providing a better delay bound, particularly at low load settings.
Through simulation results we have shown that BWAR outperforms
both traditional backpressure (at low loads) and conventional DTN-routing mechanisms
(at high loads) in encounter-based mobile networks.

There are a few open avenues for future work suggested by our study. First,  we would like to undertake a more careful analysis
of the delay improvements obtained, relating them more explicitly, for instance, to arrival process parameters and the underlying mobility model. Second, the improvements obtained by BWAR in terms of delay are obtained at the expense of greater number of transmissions due to the introduced redundancy. While this may be acceptable in some networks, for energy-constrained networks this could be a concern. We therefore plan to explore the design of energy-efficient variants of BWAR in the future, in which the redundancy can be controlled to provide a tunable tradeoff between energy and delay. We would also like to investigate automated self-configuration of the timeout parameter for duplicate removal through a distributed mechanism, as this is currently statically configured in BWAR.

\bibliographystyle{IEEEtran}

\end{document}